**330.46:519.25**

# ПРОГНОЗУВАННЯ ФІНАНСОВИХ РЯДІВ: СЕМАНТИЧНИЙ АНАЛІЗ ЕКОНОМІЧНИХ НОВИН

## К. Ю. Кононова

Кандидат економічних наук, доцент,
доцент кафедри економічної кібернетики та прикладної економіки

Харківський національний університет ім. В. Н. Каразіна
площа Свободи, 4, м. Харків, 61022, Україна
*kateryna.kononova@gmail.com*

## А. О. Дек

Магістр з прикладної економіки,
аспірант кафедри економічної кібернетики та прикладної економіки

Харківський національний університет ім. В. Н. Каразіна
площа Свободи, 4, м. Харків, 61022, Україна
*dektox@gmail.com*

У роботі запропоновано метод прогнозування фінансових часових рядів з урахуванням семантики новинних стрічок. Для семантичного аналізу економічних новин на основі словника Loughran McDonald Master Dictionary було сформовано вибірку негативних і позитивних з фінансової точки зору слів. До вибірки увійшли слова з високою частотою згадування у новинах фінансових ринків; для однокореневих слів була залишена тільки загальна частина, що дозволило охопити одним запитом кілька слів. В якості інструментарію прогнозування використовувалися нейронні мережі. Для автоматизації процесу видобування економічної інформації з новин у програмному середовищі MATLAB Simulink розроблений скрипт, який аналізує новини компанії, спираючись на сформований словник. Проведене експериментальне дослідження з різними архітектурами нейронних мереж продемонструвало високу адекватність побудованих моделей та підтвердило доцільність використання інформації з новинних стрічок для прогнозування котирувань акцій.

**Ключові слова**: *прогнозування котирувань, стрічки фінансових новин, семантичний аналіз, нейронна мережа.*

# ПРОГНОЗИРОВАНИЕ ФИНАНСОВЫХ РЯДОВ: СЕМАНТИЧЕСКИЙ АНАЛИЗ ЭКОНОМИЧЕСКИХ НОВОСТЕЙ

## Е. Ю. Кононова

Кандидат экономических наук, доцент,
доцент кафедры экономической кибернетики и прикладной экономики

Харьковский национальный университет им. В. Н. Каразина
площадь Свободы, 4, г. Харьков, 61022, Украина
*kateryna.kononova@gmail.com*

## А. О. Дек

Магистр по прикладной экономике,
аспирант кафедры экономической кибернетики
и прикладной экономики

Харьковский национальный университет им. В. Н. Каразина
площадь Свободы, 4, г. Харьков, 61022, Украина
*dektox@gmail.com*

В работе предлагается метод прогнозирования финансовых временных рядов с учетом семантики новостных лент. Для семантического анализа экономических новостей на основе словаря Loughran McDonald Master Dictionary была сформирована выборка негативных и позитивных с финансовой точки зрения слов. В выборку вошли слова с высокой частотой упоминания в новостях финансовых рынков; для однокоренных слов была оставлена только общая часть, что позволило охватить одним запросом несколько слов. В качестве инструментария прогнозирования использовались нейронные сети. Для автоматизации процесса извлечения экономической информации из новостей в программной среде MATLAB Simulink разработан скрипт, который анализирует новости компании, опираясь на сформированный словарь. Проведенное экспериментальное исследование с различными архитектурами нейронных сетей продемонстрировало высокую адекватность построенных моделей и подтвердило целесообразность использования информации из новостных лент для прогнозирования котировок акций.

**Ключевые слова**: *прогнозирование котировок, ленты финансовых новостей, семантический анализ, нейронная сеть.*

# FINANCIAL TIME SERIES FORECASTING: SEMANTIC ANALYSIS OF ECONOMIC NEWS

## Kateryna Kononova

PhD (Economic Sciences), Docent,
Associate Professor of Department
of Economic Cybernetics and Applied Economics

V. N. Karazin Kharkiv National University
4 Svobody Sq., Kharkiv, 61022, Ukraine
*kateryna.kononova@gmail.com*

## Anton Dek

Master's Degree in Applied Economics,
PhD student, Department of Economic Cybernetics and Applied Economics

V.N. Karazin Kharkiv National University
4 Svobody Sq., Kharkiv, 61022, Ukraine
*dektox@gmail.com*

The paper proposes a method of financial time series forecasting taking into account the semantics of news. For the semantic analysis of financial news the sampling of negative and positive words in economic sense was formed based on Loughran McDonald Master Dictionary. The sampling included the words with high frequency of occurrence in the news of financial markets. For single-root words it has been left only common part that allows covering few words for one request. Neural networks were chosen for modeling and forecasting. To automate the process of extracting information from the economic news a script was developed in the MATLAB Simulink programming environment, which is based on the generated sampling of positive and negative words. Experimental studies with different architectures of neural networks showed a high adequacy of constructed models and confirmed the feasibility of using information from news feeds to predict the stock prices.

**Keywords**: *stock price forecasting, financial news, semantic analysis, neural network.*

## Постановка проблеми

Прогнозування фінансових часових рядів було і залишається одною з найактуальніших задач в економіці, оскільки є необхідним елементом будь-якої інвестиційної діяльності, що полягає у вкладанні грошей з метою отримання доходу в майбутньому. Останнім часом, коли стали доступні потужні засоби збору та обробки інформації, прогнозування котирувань на фондовому ринку стає однією з найпопулярніших задач для практичного застосування методів Data Mining[1] [7].

## Аналіз останніх досліджень і публікацій

Прогнозування фінансових показників, зокрема біржових котирувань, з використанням семантичного аналізу новин раніше розглядалось лише в контексті фундаментального аналізу. Однак після піонерського дослідження Де Лонга [3] з'явилися перші спроби формалізувати цей процес, а економісти зацікавилися дослідженням ролі настроїв інвесторів на фінансових ринках. Згідно з Бейкером і Вурглером [1] «зараз навіть не постає питання, яке виникало ще десятиліття тому: чи настрої інвесторів впливають на ціну активів; зараз актуальне інше питання: як оцінити настрої інвесторів і виміряти їх вплив».

Де Лонг та ін. показали, що у разі, коли необізнані «торговці шумом»[2] засновують свої торгові рішення на основі настроїв і не схильні до ризику, зміни настрою призводять до ще більш шумної торгівлі, більшої недооцінки і надмірної нестабільності [3]. Коган та ін. [6] стверджують, що «торговці шумом» можуть викликати значний рух цін і надлишок волатильності в короткостроковій перспективі. З цими твердженнями можна сперечатися, але безумовно, настрої інвесторів серйозно впливають на ситуа-цію на ринку, вони формуються під впливом новин, отже семантичний аналіз новин може відкрити величезні можливості з прогнозування фінансових часових рядів.

---

[1] Data Mining (укр. отримання даних, інтелектуальний аналіз даних, глибинний аналіз даних) — збірна назва, що використовується для позначення сукупності методів виявлення в даних раніше невідомих, нетривіальних, практично корисних і доступних до інтерпретації знань, необхідних для прийняття рішень у різних сферах людської діяльності. Термін введений Григорієм Пятецьким-Шапіро у 1989 році.

[2] Ірраціональна торгівля або «торгівля шумом» (англ. Noise trading) — торгівля на фондовому ринку, за якої рішення про купівлю, продаж або утримання активу ірраціональні і непостійні. Наявність таких «торговців шумом» на фінансових ринках призводить до того, що ціни і рівні ризику не збігаються з очікуваними значеннями, навіть якщо всі інші трейдери є раціональними.

Розглядаючи найуспішніші спроби автоматизації семантичного аналізу економічних новин, варто згадати роботу Да та ін. «Підводячи підсумки FEARS[1]: настрої інвесторів, ірраціональна торгівля і сукупна волатильність» [2]. У статті досліджувалась волатильність ринків залежно від пошукових запитів в інтернеті, на основі чого було складено індекс FEARS. Аналізуючи дані за період з 2004 по 2008 рр. було показано, що FEARS досить точно прогнозує волатильність у наступний день. Автори використовували сервіс Google Trends [4], що надає обсяг пошукових запитів залежно від часу. В результаті аналізу ними було знайдено залежність між змінами кон'юнктури ринку (такими як волатильність на фондовій біржі, рівень безробіття, рецесія) та кількістю пошукових запитів користувачів інтернету за тими чи іншими ключовими словами[2], які було поділено на «позитивні» та «негативні»[3].

## Мета та завдання

Метою статті є розробка методу прогнозування котирування акцій на основі семантичного аналізу економічних новин, який включає такі етапи:

1) формування вибірки ключових слів для проведення семантичного аналізу фінансових новин;

2) розробка програмного скрипта для автоматизації пошуку «позитивних» і «негативних» слів у новинних стрічках;

3) формування навчальної вибірки, що включає часові ряди фінансових котирувань і відповідні кількості «позитивних» та «негативних» слів у новинах;

---

[1] FEARS — абревіатура від англ. Financial and Economic Attitudes Revealed by Search — ставлення до фінансових та економічних питань, виявлене пошуковими запитами.

[2] Для семантичного аналізу було використано словник Harvard IV-4 dictionary [5], який містить набори слів, характерні для позитивних і негативних текстів.

[3] У роботі [2] вказується, що не всі слова, відмічені в словнику Harvard IV-4 «negative», можуть бути використані для аналізу фінансових новин, тому авторами були виділені негативні у фінансовому розумінні 40 слів, а саме: bankrupt, bankruptcy, beggar, blackmail, bribe, bum, commoner, corrupt, cost, costliness, costly, debtor, default, depression, destitute, extravagant, fine, fire, gamble, hole, hustle, hustler, inflation, jobless, laid, lay, liquidate, miser, owe, poor, recession, squander, tariff, underworld, uneconomical, unemployed, unprofitable, vagabond, vagrant, waste.

4) вибір математичного інструментарію та побудова економіко-математичної моделі, що дозволяє прогнозувати котирування акцій на основі семантичного аналізу економічних новин;

5) аналіз якості та прогностичних можливостей побудованої моделі.

### Основні результати дослідження

Для прогнозування фінансових рядів на основі семантичного аналізу новин необхідно було зібрати наступну інформацію: історичні ціни акцій, обсяги торгів, фінансові новини щодо діяльності компанії, яка аналізується. Перші дві категорії доступні для скачування на таких сервісах, як Google Finance і Yahoo Finance, остання категорія — новини — доступна на тих же сервісах, однак вимагає ґрунтовнішого збору та аналізу.

Зазвичай аналіз економічних новин проводиться людиною, оскільки тексти новин зрозумілі лише людині, вони не підпорядковані чіткій машинній логіці та можуть містити фразеологізми, метафори, гіперболи, алегорії та інші художні прийоми мови. Однак аналіз новин власноруч є дуже трудомістким і тривалим процесом. Якщо говорити про процес прийняття рішень при здійсненні торгової діяльності на фондовому ринку, то швидкість аналізу стає критичною.

На першому етапі дослідження для формування вибірки ключових слів було обрано словник Loughran McDonald Master Dictionary [8]. Він містить 80 000 слів, кожне з яких описується набором характеристик, розрахованих на основі аналізу більш ніж двохсот тисяч документів. Серед характеристик для цілей нашого аналізу суттєвими є: частота використання слова в документах; частка появи аналізованого слова від загального обсягу; стандартне відхилення частоти появи в документах; кількість документів, що містять щонайменше одне входження слова; ідентифікатори категорії (наприклад, негативні, позитивні, невизначені, суперечні, модальні, стримуючі).

Розвиваючи ідеї, запропоновані в [2], для семантичного аналізу економічних новин на основі даних словника Loughran McDonald Master Dictionary нами було сформовано власну вибірку негативних і позитивних з економічної точки зору слів. Хоча словник містить 353 позитивних з фінансової точки зору слова та 2337 негативних, але всі дві з половиною тисячі слів використовувати недоцільно через трудомісткість подальшого аналізу. У вибірку для проведення нашого дослідження було відібрано високочастотні слова (частота їх використання перевищує десятитисячну долю процента від загальної кількості слів у документах). Крім того, для однокореневих слів була залишена лише загальна частина, що дозволило охопити одним запитом кілька слів. Підсумкова вибірка містить 850 негативних і 187 позитивних слів (ця пропорція ґрунтується на висновках Жи Да та ін. [2] відносно того, що настрої інвесторів більшою мірою описуються негативними з фінансової точки зору словами).

На другому етапі для автоматизації процесу аналізу новинних стрічок авторами цієї статті було створено скрипт у програмному пакеті MATLAB Simulink. Як джерело новин обрано сервіс YAHOO Finance [9]. Скрипт автоматично заходить на сайт YAHOO Finance, переходить на сторінку зі списком новин щодо обраної компанії, виділяє в HTML-коді потрібний блок з новинами, потім знаходить усі посилання на цій сторінці, що ведуть на новини, групує їх за датами, заходить на кожне з посилань, шукає негативні та позитивні з фінансової точки зору слова, заносить дані про кількість таких слів у масив даних і зберігає їх до файлу в форматі .csv. Потім скрипт переходить до більш ранніх новин за посиланням «Older Headlines» і повторює вищеописану процедуру.

У якості прикладу для прогнозування котирувань акцій було обрано компанію Royal Dutch Shell (RDS-A), для якої було сформовано вибірку, що складається з двохсот елементів і містить такі дані: котирування (ціна закриття), об'єм торгів, кількість позитивних, негативних слів з фінансової точки зору. Математичним інструментарієм у дослідженні було обрано нейронні мережі, оскільки вони не накладають суттєвих обмежень на характер вхідної інформації та для їх побудови немає потреби в апріорних гіпотезах.

На основі сформованої вибірки даних для прогнозування котирувань було створено та проведено навчання декількох архітектур нейронних мереж прямої передачі сигналу, що відрізнялися як структурою подання та обробки вхідної інформації, так і кількістю та розмірністю прихованих шарів (фрагмент експериментальних розрахунків наведено в табл. 1). Проведення навчання нейронних мереж і тестове прогнозування здійснювалось на основі таких вхідних показників:
- $c$ — котирування (ціна закриття),
- $c\tau$ — котирування з лагом $\tau$, $\tau = 1, …, 3$,
- $v$ — об'єм торгів,
- $p$ — кількість позитивних слів,
- $n$ — кількість негативних слів,
- $d$ — дата.

*Таблиця 1*

**АРХІТЕКТУРИ НЕЙРОМЕРЕЖ, ЯКІ БУЛИ ДОСЛІДЖЕНІ**

| № | Вхід | Архітектура | Похибка навчання, MSE | Відносна похибка прогнозування | Якість прогнозування, adjusted $R^2$ |
|---|------|-------------|-----------------------|-------------------------------|--------------------------------------|
| 1 | $p, n$ | 2-1 | 4,9047 % | 19,28 % | 97,188 % |
| 2 | $p, n, c1$ | 3-1 | 0,2138 % | 1,04 % | 99,497 % |
| 3 | $p, n, c1, c2$ | 4-2-1 | 0,2128 % | 1,35 % | 99,956 % |
| 4 | $p, n, c1, c2, c3$ | 5-3-1 | 0,1825 % | **0,55 %** | 99,955 % |
| 5 | $p, n, c1, c2, c3, d$ | 6-3-1 | 0,1639 % | 0,88 % | **99,959 %** |
| 6 | $p, n, c1, c2, c3, d, v$ | 7-3-1 | **0,1236 %** | 3,92 % | 99,802 % |
| 7 | $p/n, c1, c2, d$ | 4-2-1 | 0,1715 % | 2,10 % | 99,925 % |
| 8 | $c1, c2, c3, d$ | 4-3-1 | 0,1929 % | 4,28 % | 99,710 % |

Дослідження якості навчання та прогностичних властивостей моделей дало можливість визначити три найбільш адекватні нейронні мережі (четверта, п'ята та шоста), що дає підстави обирати одну із них для прогнозування розвитку фінансових показників у реальних умовах. Наприклад, четверта мережа складається із вхідного, одного прихованого та вихідного шарів: на вхід мережі подається ціна закриття за три попередніх дня та кількості позитивних і негативних слів протягом сьогоднішнього дня; виходом мережі є ціна закриття на сьогодні (рис. 1).

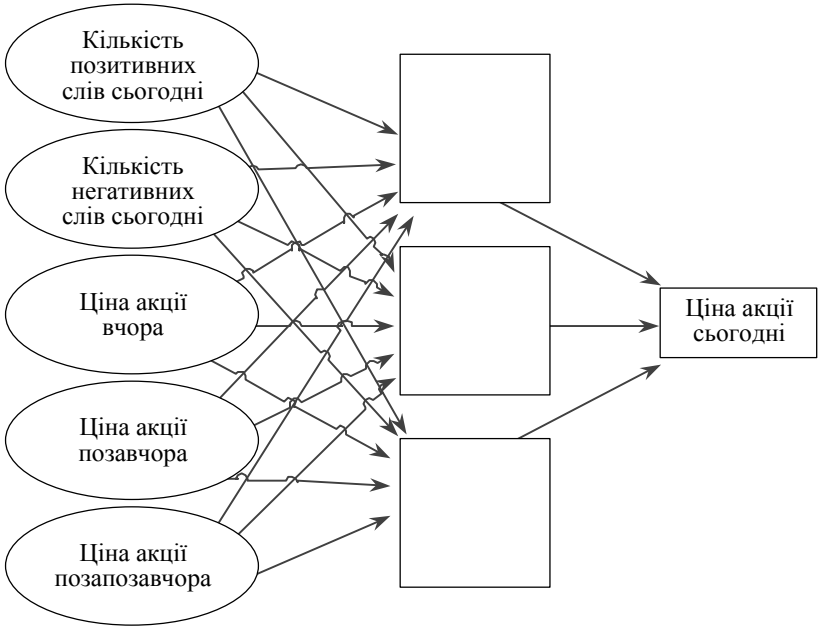

Рис. 1. Архітектура нейронної мережі

Як можна бачити з рис. 2, для цієї моделі середня квадратична похибка моделювання навчального масиву даних з двохсот елементів не перевищує 0,18 %.

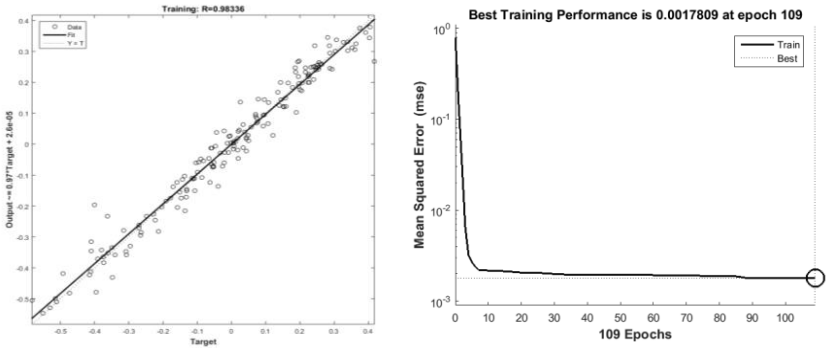

Рис. 2. Діаграма розсіювання та динаміка похибки навчання

Четверта модель зі структурою 5-3-1 дозволила доволі точно описати історичний ряд котирувань (перші 175 елементів на рис. 3) і масив даних, використаних для оцінки якості прогнозування (див. фрагмент із 176 по 195 моменти часу на рис. 3).

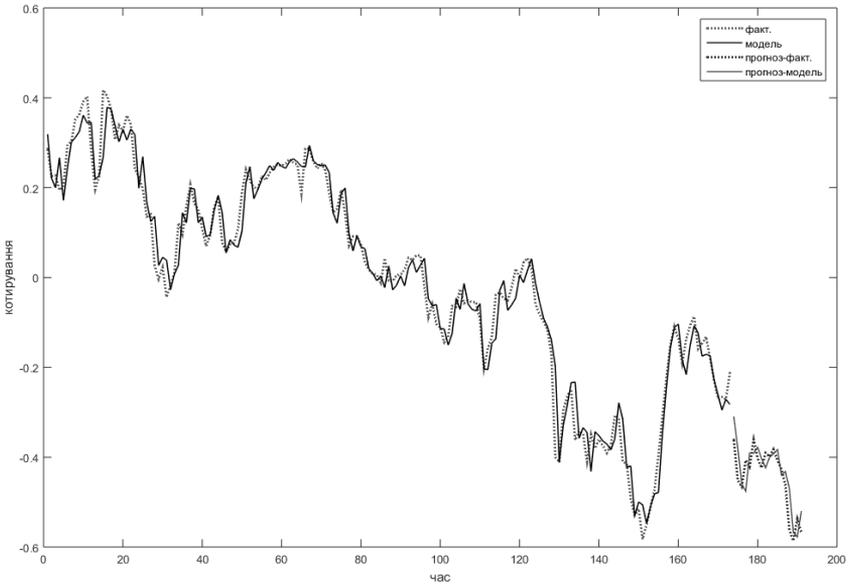

Рис. 3. Ряди фактичних і змодельованих цін акцій Royal Dutch Shell з урахуванням семантики новин (четверта мережа)

Восьма нейронна мережа зі структурою 4-3-1 продемонструвала результати моделювання часового ряду цін акцій Royal Dutch Shell на навчальній (перші 175 елементів на рис. 4) і тестовій (від 176 по 195 моменти часу на рис. 4) вибірках, як зображено нижче.

Як можна бачити з рис. 3 та табл. 1, прогностичні властивості моделей, що враховують кількості позитивних і негативних з фінансової точки зору новин, є вищими порівняно з моделями, у яких семантика новин не враховується. Відповідно, результати проведених модельних експериментів обумовлюють доцільність використання інформації з новинних стрічок для прогнозування котирувань акцій.

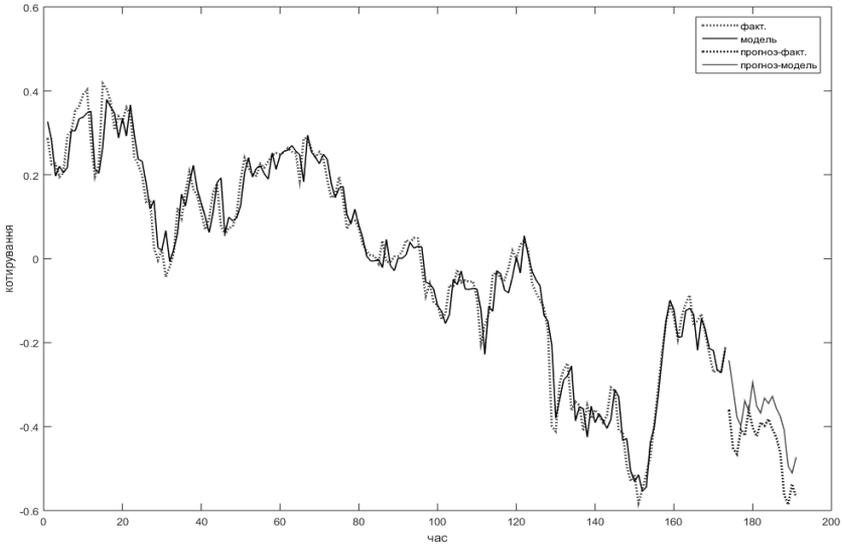

Рис. 4. Ряди фактичних і змодельованих цін акцій Royal Dutch Shell
без урахування семантики новин (восьма мережа)

### Висновки

У роботі запропоновано метод моделювання фінансових часових рядів з урахуванням семантики новинних стрічок. Основу методу склали: 1) алгоритм формування вибірки високочастотних негативних і позитивних з фінансової точки зору слів; 2) скрипт, розроблений у програмному середовищі MATLAB Simulink, для автоматизація процесу вилучення економічної інформації з новин, спираючись на сформований словник; 3) математичним інструментарієм прогнозування було обрано нейронні мережі. Порівняльний аналіз різних архітектур нейронних мереж дозволив обґрунтувати доцільність використання семантики новинних стрічок для прогнозування котирувань акцій та продемонстрував достатньо високу адекватність побудованих моделей.

### Литература

1. *Baker M.* Investor Sentiment and the Cross-Section of Stock Returns / M. Baker, J. Wurgler // The Journal of Finance. — 2006. — Vol. LXI. — № 4. — P. 1645—1680.

2. *Da Z.* The Sum of All FEARS: Investor Sentiment, Noise Trading and Aggregate Volatility [Електронний ресурс] / Z. Da, J. Engelberg, P. Gaox. — Режим доступу : http://rady.ucsd.edu/faculty/directory/engelberg/pub/portfolios/FEARS.pdf.

3. *DeLong S.S.W.* Noise Trader Risk in Financial Markets // The Journal of Political Economy. — 1990. — Vol. 98. — № 4. — P. 703-738.

4. Google Trends [Електронний ресурс]. — Режим доступу : https://www.google.com/trends/.

5. Harvard IV-4 dictionary [Електронний ресурс]. — Режим доступу : http://www.wjh.harvard.edu/~inquirer/homecat.htm.

6. *Kogan L.* Price Impact and Survival of Irrational Traders [Електронний ресурс] / L. Kogan, S. Ross, J. Wang, M. Westerfield // FAME Research Paper Series. — Research Paper № 116. — International Center for Financial Asset Management and Engineering. — Режим доступу : http://www.swissfinanceinstitute.ch/rp116.pdf.

7. *Kovalerchuk B.* Data mining in finance: advances in relational and hybrid methods / B. Kovalerchuk, E. Vityaev. — Norwell : Kluwer Academic Publishers, 2000. — 308 p.

8. *McDonald B.* Loughran and McDonald Sentiment Word Lists [Електронний ресурс] / B. McDonald. — Режим доступу : http://www3.nd.edu/~mcdonald/Word_Lists.html.

9. YAHOO Finance [Електронний ресурс]. — Режим доступу : http://finance.yahoo.com/.

**References**

1. Baker, M., & Wurgler, J. (2006). Investor Sentiment and the Cross-Section of Stock Returns. *The Journal of Finance, 61*(4), 1645—1680.

2. Da, Z., Engelberg, J., & Gaox, P. (2009). *The Sum of All FEARS: Investor Sentiment, Noise Trading and Aggregate Volatility*. Retrieved from http://rady.ucsd.edu/faculty/directory/engelberg/pub/portfolios/FEARS.pdf.

3. DeLong, S.S.W. (1990). Noise Trader Risk in Financial Markets.*The Journal of Political Economy, 98*(*4*), 703—738.

4. Google Trends. (2015, September 30). Retrieved from https://www.google.com/trends/.

5. Harvard IV-4 dictionary. (2015, November 2). Retrieved from http://www.wjh.harvard.edu/~inquirer/homecat.htm.

6. Kogan, L., Ross, S., Wang, J., & Westerfield, M. (2004). Price Impact and Survival of Irrational Traders. *FAME Research Paper Series, 116.* Retrieved from http://www.swissfinanceinstitute.ch/rp116.pdf.

7. Kovalerchuk, B., & Vityaev, E. (2000). *Data mining in finance: advances in relational and hybrid methods.* Norwell: Kluwer Academic Publishers.

8. Loughran and McDonald Sentiment Word Lists. (2015, November 8). Retrieved from http://www3.nd.edu/~mcdonald/Word_Lists.html.

9. YAHOO Finance. (2016, January 6). Retrieved from http://finance.yahoo.com/.